\documentclass[journal]{IEEEtran}

\usepackage{cite}
\usepackage{amsmath,amssymb,amsfonts,bm}
\usepackage{amsthm}
\usepackage{graphicx}
\usepackage{textcomp}
\usepackage{xcolor}
\usepackage{subcaption}
\usepackage{enumitem}
\usepackage{placeins}
\usepackage{url}

\usepackage[ruled,vlined]{algorithm2e}
\SetKw{KwDownTo}{downto}

\begin{document}

\title{A hidden Markov model for unsupervised EEG-based tracking of auditory attention\\ to a single voice}

\author{
Tianyi Li$^{1}$, Simon Geirnaert$^{1,3}$, Davina Van den Broek$^{2}$, Elien Bellon$^{2}$, Bert De Smedt$^{2}$ and Alexander Bertrand$^{1}$%
\thanks{*This project has received funding from Internal Funds KU Leuven
(projects IDN/23/006, C14/25/108, and C3/25/107), the FWO
(Research Foundation Flanders) for projects G081722N and G026026N,
a Junior and a Senior Postdoctoral Fellowship for Fundamental Research
awarded to S. Geirnaert (No. 1242524N and 1211827N),
and a Junior Postdoctoral Fellowship for Fundamental Research
awarded to E. Bellon (No. 12C9523N).}%
\thanks{$^{1}$T. Li, S. Geirnaert, and A. Bertrand are with the Department of Electrical Engineering (ESAT),  STADIUS Center for Dynamical Systems, Signal Processing
and Data Analytics and with Leuven.AI – KU Leuven institute for AI.}%
\thanks{$^{2}$D. Van den Broek, E. Bellon, and B. De Smedt are with the Faculty of Psychology and Educational Sciences, KU Leuven, 3000 Leuven, Belgium.}

\thanks{$^{3}$S. Geirnaert is also with KU Leuven, Department of Neurosciences,
Research Group ExpORL.}
}


\maketitle

\begin{abstract}
\textbf{Objective:}
Absolute auditory attention decoding (aAAD) aims to track how strongly a listener attends to a single speech source using electroencephalography (EEG). Unlike selective auditory attention decoding, aAAD lacks a competing speech reference, making reliable attention tracking challenging.

\textbf{Approach:}
We propose a fully unsupervised aAAD pipeline combining neural envelope tracking (NET) with hidden Markov model (HMM)-based temporal inference. Using EEG signals and the target speech envelope, the framework estimates a continuous attention trajectory without explicit attention-state labels. We evaluate the approach using EEG recordings from 30 adults, investigate the impact of analysis window length, and compare HMM post-processing with moving-average (MA) smoothing and alternative EEG features.

\textbf{Main results:}
With 1-s analysis windows, non-causal HMM post-processing improves the area under the receiver operating characteristic curve (AUC) for all 30 participants, increasing mean AUC from 0.547 for raw NET to 0.832. A 5-s window achieves the highest mean AUC of 0.858. Compared with non-causal MA smoothing, the HMM achieves higher AUC across the examined smoothing scales. Individual switch detection times are comparable, while the HMM generally achieves lower group-level detection times. NET also outperforms alpha power and spectral entropy as an observation feature in the unsupervised HMM framework.

\textbf{Significance:}
The proposed framework provides a label-free approach for estimating temporal fluctuations in auditory attention, demonstrating the utility of probabilistic temporal inference for continuous attention tracking in single-speaker listening scenarios.
\end{abstract}


\IEEEpeerreviewmaketitle

\section{Introduction}
Auditory attention plays a crucial role in contexts such as classroom environments, where the instructor's voice often acts as the primary attractor. However, attention is not a static cognitive state; instead, it fluctuates over time and is influenced by factors such as distractions, fatigue \cite{warm2008vigilance}, and mind wandering \cite{smallwood2015science}. Understanding how auditory attention evolves over time can therefore provide valuable insights into learning processes in the classroom and enable applications such as attention monitoring systems that track students' engagement dynamics and provide student-specific feedback \cite{poulsen2017eeg}.

In the past decade, extensive research has demonstrated that auditory attention to natural speech can be decoded from EEG signals with stimulus reconstruction approaches \cite{geirnaert2021electroencephalography,o2015attentional,etard2019decoding,kubanek2013tracking}.
Compared to commonly used EEG markers associated with attention such as alpha power \cite{KLIMESCH199873,benedek2014alpha} and spectral entropy \cite{lesenfants2020interplay}, neural envelope tracking (NET) has been regarded as a stimulus-specific metric of auditory attention, as it reflects neural responses to the speech signal itself rather than general mental effort \cite{roebben2024you,lesenfants2020interplay,vanthornhout2019effect}. To compute the NET metric, a neural decoder is trained to reconstruct the temporal envelope of the attended speech from EEG recordings, and the Pearson correlation between the reconstructed envelope and the corresponding speech envelope is used as a measure of auditory attention. This NET-based approach has been successfully applied in selective auditory attention decoding (sAAD) \cite{geirnaert2021electroencephalography, etard2019decoding, o2015attentional} and absolute auditory attention decoding (aAAD) paradigms \cite{roebben2024you, EMBC, vandenbroek_bellon_li_geirnaert_bertrand_desmedt_2025,levy2025differential}.

In sAAD, attention is measured by comparing correlations across multiple competing speakers, with the speaker yielding the highest correlation identified as the attended one. In contrast, aAAD considers scenarios in which only a single speech stream is present and aims to continuously track how attention to this target evolves over time. It directly uses the NET score as a measure of the degree of attention to the target speaker. However, as there is no relative reference from competing speakers to compare with, aAAD poses a fundamentally more challenging inference problem than sAAD, as attention must be estimated from a single, neural evidence stream.

Moreover, instantaneous NET estimates are typically noisy, in particular when measured over short time intervals. They can also be influenced by confounding factors that are not attention-driven, such as temporal variations in the dynamic range of the speech envelope \cite{EMBC, haupt2025neural}. As a result, raw NET values can be difficult to interpret directly on a window-by-window basis.
 Motivated by these limitations, it is therefore natural to post-process the NET score by treating the NET sequence as a time series and applying smoothing techniques that take temporal coherence into account. This is reasonable because attention states in neighboring windows are not expected to be independent, especially when the window length is short. Instead, attention is likely to be piecewise stable in natural learning scenarios, as people are unlikely to switch attention every second and tend to remain focused on the target for at least several seconds, or even minutes \cite{langner2013sustaining}.

In this work, we present a modular aAAD pipeline that combines NET-based evidence extraction with probabilistic temporal inference based on a hidden Markov model (HMM), a powerful framework that is able to infer attention states from NET observations, inspired by similar work in sAAD \cite{heintz2025postprocessingeegbasedauditoryattention}. The proposed pipeline is designed for off-line analysis and is fully unsupervised (i.e., no attention labels are used for training), relying only on a small set of tunable and interpretable parameters.

The remainder of this paper is organized as follows. Section~\ref{sec:methodology} introduces the proposed aAAD pipeline, including the NET-based evidence extraction module and the HMM-based temporal inference module. Section~\ref{sec_exp} describes the experimental protocol, EEG preprocessing procedure, evaluation metrics, and default parameter settings. Section~\ref{sec:results} presents the experimental validation of the proposed framework. We first demonstrate the added value of HMM post-processing compared with raw NET, and then analyze the effect of analysis window length on the area under the curve (AUC) and switch detection time (SDT). We further compare the HMM with moving-average (MA) smoothing to assess the advantage of state-based temporal inference over fixed-window averaging. Finally, we compare NET with alpha power and spectral entropy as observation features for the unsupervised HMM framework. Section~\ref{sec_conc} concludes the paper.

\section{Methodology}
\label{sec:methodology}

\subsection{Pipeline Overview}\label{sec_pipeline}
\begin{figure*}[t]
  \centering
  \includegraphics[
      width=\textwidth,
      trim=0 150 0 120,  
      clip
  ]{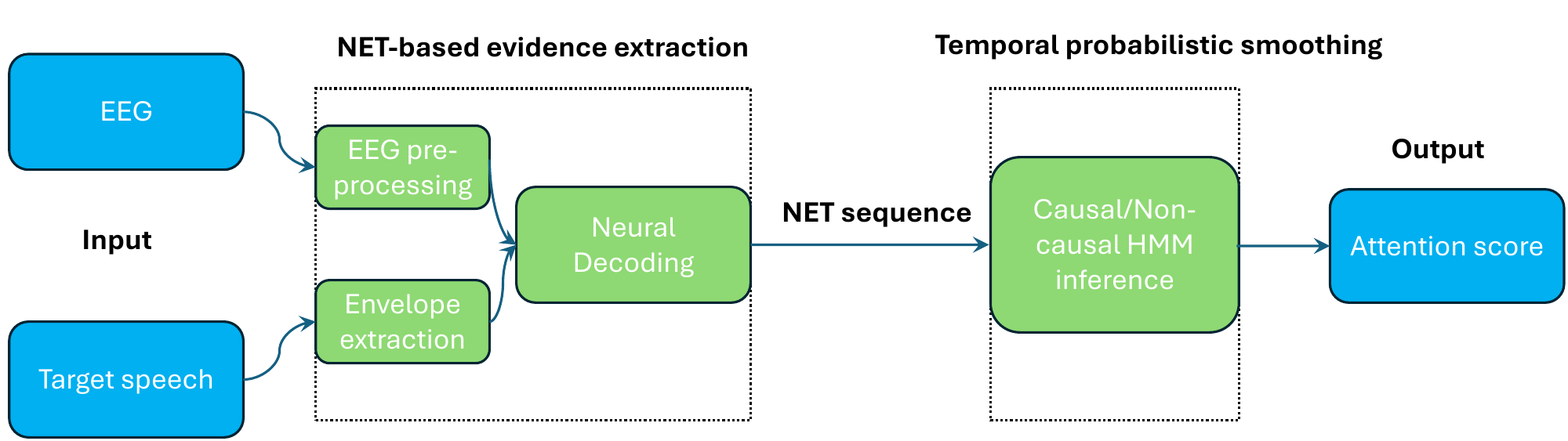}
  \caption{Overview of the proposed aAAD pipeline.}
  \label{fig:pipeline}
\end{figure*}

This section presents an overview of the proposed pipeline for off-line analysis of absolute auditory attention, which is conceptually visualized in Fig. \ref{fig:pipeline}. The pipeline operates in a fully unsupervised manner and is controlled by a small set of interpretable parameters.

Starting from the input EEG recordings and the target speech signal, the proposed pipeline first performs EEG pre-processing and speech envelope extraction, followed by neural decoding to compute a windowed neural envelope tracking (NET) sequence, which serves as stimulus-specific evidence of attention to the target speech. Concretely, a linear spatio-temporal decoder is first trained to extract the neural responses to the target speech, thereby removing irrelevant EEG signal components. To this end, we use the so-called stimulus-reconstruction paradigm, which aims to reconstruct the speech envelope from the pre-processed EEG by solving a least-squares regression problem on all data. The trained decoder is then applied to the EEG to obtain a reconstructed envelope, which is compared with the target speech envelope using the Pearson correlation coefficient computed over multiple non-overlapping\footnote{We use non-overlapping windows, as the HMM post-processing assumes independent observations.} windows. The resulting correlation values form the NET sequence. Although a linear least-squares decoder is used in this work, the proposed pipeline is not restricted to a specific decoding method, and any alternative linear or nonlinear decoder (e.g., CCA-based \cite{de2018decoding} or neural network-based models \cite{accou2023decoding,xu2024convconcatnet}) can be incorporated, as long as it provides a correlation-based neural tracking metric between EEG and the target stimulus. Next, an HMM is applied to the NET sequence—using either causal or non-causal inference—to explicitly model temporal dependencies and obtain a smoothed attention estimate in the form of the posterior probability of being attentive. This posterior probability is finally reported as the continuous attention score over time. Importantly, both the neural evidence extraction and the temporal inference stages operate without requiring attention-state labels.

\subsection{Evidence Extraction Module: Neural Envelope Tracking}
\label{sec_net}

\subsubsection{Neural Decoder Design}
Similar to \cite{geirnaert2021electroencephalography,etard2019decoding,o2015attentional,roebben2024you,EMBC}, we adopt a spatio-temporal linear decoder that aims to reconstruct the speech envelope of the presented stimulus from the EEG of the participant. In this work, the speech envelope is extracted using the power law and gammatone filterbank procedure proposed in \cite{biesmans2016auditory}, yet any envelope extraction method can be used. Formally, we train a decoder $\hat{\mathbf{d}} \in \mathbb{R}^{LC \times 1}$ over $K$ time training samples using all data such that \cite{biesmans2016auditory}:
\begin{equation}\label{eq_lsq}
\hat{\mathbf{d}}=\underset{\mathbf{d}}{\operatorname{argmin}}\left\|\mathbf{s}-\mathbf{X d}\right\|_2^2,
\end{equation}
where $\mathbf{s}=[s(0) \cdots s(K-1)]^\top\in \mathbb{R}^{K \times 1}$ denotes the $K$ training samples of the speech envelope, $\mathbf{X}=[\mathbf{X}_1 \cdots \mathbf{X}_C] \in \mathbb{R}^{K \times LC}$ collects $L$ time-lagged copies of each EEG channel ($C$ channels in total), where $\mathbf{X}_c$ denotes the data for channel $c$ such that
\begin{equation}
\mathbf{X}_c =
\begin{bmatrix}
x_c(0) & x_c(1) & \cdots & x_c(L-1) \\
x_c(1) & x_c(2) & \cdots & x_c(L) \\
\vdots & \vdots & \ddots & \vdots \\
x_c(K-1) & 0 & \cdots & 0
\end{bmatrix}.
\end{equation}
A time delay of 0 to 500 ms \cite{vanthornhout2019effect} is chosen corresponding to $L=9$ at the 16 Hz sampling rate (see Section \ref{data_acq}). The closed-form solution for this least-squares problem \eqref{eq_lsq} is given by
\begin{equation}\label{eq_lsq_sol}
\hat{\mathbf{d}}=(\mathbf{X}^\top \mathbf{X})^{-1} \mathbf{X}^\top \mathbf{s}.
\end{equation}

This decoder is trained per participant separately over the data from the entire recording ($K$ samples in total). After training, the decoder is applied to the same segments of the same participant, which are segmented into windows indexed by $t$. For each window $t$, the EEG data is represented as $\mathbf{X}^{(t)} \in \mathbb{R}^{N \times LC}$, where $N$ denotes the number of time samples within that analysis window. Note that $t$ is not used as a sample index, but as the window index. The decoder generates $\hat{\mathbf{s}}^{(t)} = \mathbf{X}^{(t)} \hat{\mathbf{d}}$, which represents a reconstruction of the true speech envelope $\mathbf{s}^{(t)}$ within window $t$. We define the Pearson correlation coefficient between the reconstructed and true speech envelope $\rho\left(\hat{\mathbf{s}}^{(t)}, \mathbf{s}^{(t)}\right)$ as the neural envelope tracking (NET) metric for window $t$ \cite{roebben2024you}. It was demonstrated in \cite{roebben2024you,vanthornhout2019effect,EMBC,vandenbroek_bellon_li_geirnaert_bertrand_desmedt_2025} that higher NET values on average correspond to a higher level of attention to the target speech.

\subsubsection{Label-Free Decoder Training}
In this work, the neural decoder is trained using all available data and subsequently applied to the same dataset to compute the Pearson correlation between the reconstructed and the true speech envelope on a window-by-window basis. This validation strategy differs from many previous studies \cite{biesmans2016auditory, accou2023decoding, heintz2025postprocessingeegbasedauditoryattention,yao2024identifying}, where a strict train-test split is enforced. Nevertheless, we argue that it is appropriate for the present aAAD setting, as previous work has demonstrated the feasibility of unsupervised aAAD without a separate training dataset \cite{heintz2025unsupervised}.

Specifically, the decoder is designed to capture the stimulus-response relationship between EEG signals and the speech envelope, rather than to explicitly discriminate between attentive and inattentive states. Under this formulation, inattentive EEG does not correspond to a distinct neural pattern, but rather to a weaker or noisier manifestation of the same stimulus-driven neural response. Accordingly, we adopt the perspective that aAAD is better formulated as a signal matching problem with attention-dependent signal-to-noise ratios, rather than as a binary classification problem with separable EEG patterns.

Within this framework, correlation-based evaluation provides a meaningful measure of neural tracking strength, even when the decoder is applied to the same data used for training. This formulation constitutes one of the key ingredients that enables a fully unsupervised aAAD framework, as it allows stimulus-specific neural evidence to be extracted without requiring explicit attention labels.

It should be noted that the training of the neural decoder requires sufficient baseline data during which the participant actually attends to the speech, in order to reliably estimate the stimulus–response mapping. If the available baseline segments are too limited in duration, the estimated decoder may become unstable, which can subsequently affect the quality of the neural tracking feature used for temporal inference.

\subsection{Temporal Inference Module}
\label{sec_hmm}

Given a NET sequence (containing one NET value for each window $t$), we employ a hidden Markov model (HMM) as a temporal inference
module to obtain a smooth estimate of the underlying attention dynamics, inspired by similar work in sAAD \cite{heintz2025postprocessingeegbasedauditoryattention}.
The HMM takes the NET sequence as its sole observation input and infers a latent attention
state sequence that evolves over time. We note that the temporal inference module
operates independently of how the NET sequence is obtained. Although NET is computed in this work
using a linear neural decoding model, the HMM can in principle be applied to any stimulus-specific
evidence sequence derived from linear or non-linear models \cite{heintz2025postprocessingeegbasedauditoryattention}.


\subsubsection{Markov Chain Background}
\begin{figure}[t]
  \centering
  \includegraphics[
      width=1.1\columnwidth,   
      trim=30 40 0 40,          
      clip
  ]{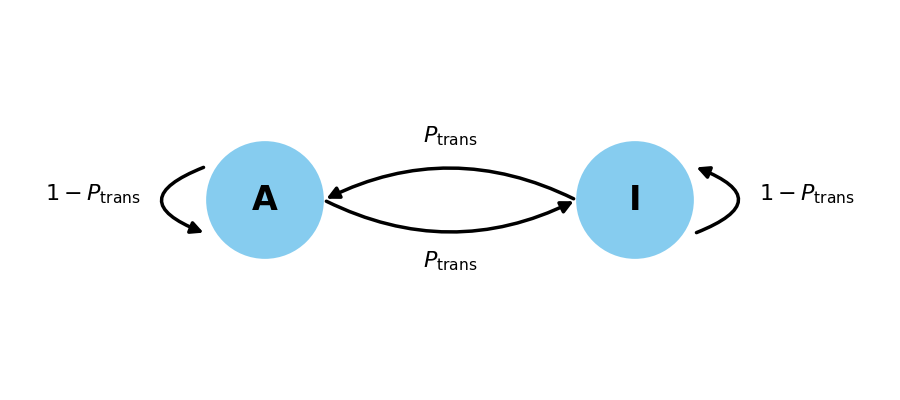}
  \caption{Two-state Markov chain used to model attention-state persistence and transitioning.}
  \label{fig:Markov}
\end{figure}

We model the temporal evolution of auditory attention using a discrete-time Markov chain with two latent states, $S_1$ (attentive) and $S_2$ (inattentive). Let $S(t)\in\{S_1,S_2\}$ denote the latent state at time index $t$ (corresponding to one window in the evidence extraction module described in Subsection \ref{sec_net}). In Fig.~\ref{fig:Markov}, these states $S_1$ and $S_2$ are labeled as A and I for readability.

The state dynamics are governed by transition probabilities, which act as hyperparameters controlling the expected dwell time of each attention state and, consequently, the temporal smoothness of the inferred attention trajectory. For simplicity and interpretability, we assume symmetric transitions between the two states:
the probability of transitioning from one state to the other is $P_{\mathrm{trans}}$, while the
self-transition probability equals $1-P_{\mathrm{trans}}$.
Equivalently, the transition matrix $\mathbf{A}$ can be written as
\begin{equation}
\mathbf{A}
=
\begin{bmatrix}
1-P_{\mathrm{trans}} & P_{\mathrm{trans}}\\
P_{\mathrm{trans}}   & 1-P_{\mathrm{trans}}
\end{bmatrix},
\label{eq:A_sym}
\end{equation}
where $A_{ij}=p(S(t)=S_j \mid S(t-1)=S_i)$.

In the context of aAAD, attention is expected to be temporally persistent, i.e., listeners are more likely to remain in the same attentional state across consecutive windows than to change states frequently.
This is captured by choosing a small value for $P_{\mathrm{trans}}$.
In summary, the Markov chain models the temporal dynamics of the latent attention state. To connect this latent process to the observed neural evidence, i.e., the windowed NET sequence, we next extend it to an HMM.

\subsubsection{Hidden Markov Model for aAAD}
Building on the Markov chain model above, we formulate an HMM by introducing a probabilistic
relationship between the latent attention states and the observed NET sequence.
This modeling choice reflects two key characteristics of attentional dynamics: (1) attentional engagement exhibits temporal persistence rather than arbitrary rapid transitions, and (2) the distribution of NET observations is state-dependent. Accordingly, the formulation adopted in this work is based on two standard HMM assumptions.

\textbf{Assumption 1 (Markovian state dynamics):}
The latent attention state at time $t$ depends only on the state at time $t-1$:
\begin{equation}
p\big(S(t)\mid S(0\!:\!t-1)\big)
=
p\big(S(t)\mid S(t-1)\big),
\label{eq:markov_assumption}
\end{equation}
where $x(i\!:\!j)$ denotes the entire sequence $\{x(i), x(i+1), \cdots, x(j)\}$.

\textbf{Assumption 2 (Conditional independence of observations):}
The NET observation $y(t)$ at time $t$ (i.e., window $t$) depends only on the current latent state:
\begin{equation}
p\big(y(t)\mid S(0\!:\!t),\,y(0\!:\!t-1)\big)
=
p\big(y(t)\mid S(t)\big),
\label{eq:emission_CI}
\end{equation}
where $y(t) \in \mathbb{R}$ denotes the NET value (or a transformed version thereof) at time index $t$.

We emphasize that Assumptions~1 and~2 are modeling simplifications rather than
literal descriptions of the underlying neural mechanisms. Assumption~1 imposes
a first-order Markov structure on the latent attention states, meaning that the
current state depends only on the immediately preceding state. This limits the
temporal memory of the model and enables efficient inference, while still
allowing attention states to persist through appropriate transition
probabilities. Assumption~2 represents the observation process using a
state-dependent model, which provides a compact description of the relationship
between attention states and the NET sequence. Although these assumptions may
not always hold exactly in practice, they offer an interpretable, computationally efficient, and effective framework for
temporal inference. All subsequent derivations are based on these two assumptions together with standard probability rules, including the chain rule and marginalization. For simplicity, intermediate algebraic steps are omitted, and we refer to \cite{heintz2025postprocessingeegbasedauditoryattention} for more details.

\subsubsection{Inference Objectives: Off-line and On-line Posteriors}
The goal of temporal inference is to estimate the latent attention state at each time point.
Two posterior quantities are of particular interest: the \textbf{non-causal posterior}
\begin{equation}
p\big(S(t)\mid y(0\!:\!T)\big),
\label{eq:posterior_non-causal}
\end{equation}
which uses the full observation sequence and is appropriate for off-line analysis, and the \textbf{causal posterior}
\begin{equation}
p\big(S(t)\mid y(0\!:\!t)\big),
\label{eq:posterior_causal}
\end{equation}
which uses only past and current observations and supports causal, real-time tracking.

\subsubsection{Inference Algorithms}
We next describe the standard HMM inference procedures used in this work.
We first present the recursions in the probability domain for clarity, and then
provide a numerically stable log-domain implementation based on the log-sum-exp trick \cite{heintz2025postprocessingeegbasedauditoryattention}.

\paragraph{Basic quantities}
Let $\pi_i \triangleq p(S(0)=S_i)$ denote the initial state probabilities and
\begin{equation}
b_i(t)
\triangleq
p\big(y(t)\mid S(t)=S_i\big),
\label{eq:bi_def}
\end{equation}
denote the emission likelihood at time $t$ for state $S_i$, for which the probability distribution estimation procedure will be explained further on (see 'Emission model').
The transition probabilities are collected in $\mathbf{A}$, as shown in \eqref{eq:A_sym}. 

\paragraph{Forward-backward recursions}
Define the forward message
\begin{equation}
\alpha_i(t)
\triangleq
p\big(S(t)=S_i,\,y(0\!:\!t)\big),
\label{eq:alpha_def2}
\end{equation}
and the backward message
\begin{equation}
\beta_i(t)
\triangleq
p\big(y(t\!+\!1\!:\!T)\mid S(t)=S_i\big),
\label{eq:beta_def2}
\end{equation}
which are initialized as
\begin{equation}
\alpha_i(0)
=
\pi_i\,b_i(0),
\label{eq:alpha_init2}
\end{equation}
and
\begin{equation}
\beta_i(T)
=
1, \forall i.
\label{eq:beta_init2}
\end{equation}

Using Assumptions~1-2 together with standard probability rules (chain rule and marginalization),
the forward recursion for $t\ge 1$ is given by \cite{heintz2025postprocessingeegbasedauditoryattention}
\begin{equation}
\alpha_i(t)
=
b_i(t)
\sum_{j\in\{1,2\}}
A_{ji}\,\alpha_{j}(t-1).
\label{eq:alpha_rec2}
\end{equation}

Similarly, the backward recursion for $t\le T-1$ follows as
\begin{equation}
\beta_i(t)
=
\sum_{j\in\{1,2\}}
A_{ij}\,b_j(t\!+\!1)\,\beta_{j}(t+1).
\label{eq:beta_rec2}
\end{equation}

From the forward messages, the causal posterior is obtained by normalization:
\begin{equation}
p\big(S(t)=S_i \mid y(0\!:\!t)\big)
=
\frac{\alpha_i(t)}{\sum_{k\in\{1,2\}} \alpha_k(t)}.
\label{eq:post_from_alpha2}
\end{equation}

Combining the forward and backward messages yields the non-causal posterior:
\begin{equation}
p\big(S(t)=S_i \mid y(0\!:\!T)\big)
=
\frac{\alpha_i(t)\,\beta_i(t)}{\sum_{k\in\{1,2\}}\alpha_k(t)\,\beta_k(t)}.
\label{eq:post_from_alpha_beta2}
\end{equation}

\paragraph{Log-domain implementation}

At this point, all components required for forward and forward-backward inference
have been specified. However, direct evaluation of the probability-domain recursions
in \eqref{eq:alpha_rec2}-\eqref{eq:beta_rec2} involves repeated multiplications of
small probabilities and may therefore suffer from numerical underflow when the
sequence length $T$ is large. To ensure numerical stability, all computations are
performed in the log-domain.

We define the log-messages
\begin{equation}
\tilde{\alpha}_i(t) \triangleq \log \alpha_i(t), \qquad
\tilde{\beta}_i(t) \triangleq \log \beta_i(t).
\end{equation}

The initialization becomes
\begin{equation}
\tilde{\alpha}_i(0)
=
\log \pi_i
+
\log b_i(0),
\label{eq:log_alpha_init}
\end{equation}
and
\begin{equation}
\tilde{\beta}_i(T)
=
0,
\label{eq:log_beta_init}
\end{equation}
where the latter follows from $\beta_i(T)=1$.

To evaluate sums of exponentials in the log-domain, we use the standard log-sum-exp
operation
\begin{equation}
\log\Big(\sum_k e^{\ell_k}\Big)
=
m + \log\Big(\sum_k e^{\ell_k-m}\Big),
\label{eq:lse}
\end{equation}
with $m=\max_k \ell_k$, which prevents numerical overflow.

Algorithm \ref{alg:log_fb} summarizes the full inference algorithm.

\begin{algorithm}[t]
\caption{Log-domain forward and forward-backward inference.
This algorithm is non-causal. For causal inference, skip the backward pass
and set $\tilde{\beta}_i(t)=0$ for all $t,i$.}\label{alg:log_fb}

\KwIn{Transition probabilities $A_{ij}$, emission probabilities $b_i(t)$, initial distribution $\pi_i$}
\KwOut{$p(S(t)=S_i \mid y(0{:}T))$, $\forall t\in[0,T]$}

\BlankLine
\textbf{Initialize} $\forall i\in\{1,2\}$:\;
$\tilde{\alpha}_i(0)\leftarrow \log\pi_i+\log b_i(0)$\;
$\tilde{\beta}_i(T)\leftarrow 0$\;

\BlankLine
\textbf{Forward pass:}\;
\For{$t\gets 1$ \KwTo $T$}{
  \For{$i\gets 1$ \KwTo $2$}{
    \For{$j\gets 1$ \KwTo $2$}{
      $\ell(j)\leftarrow \tilde{\alpha}_j(t-1)+\log A_{ji}+\log b_i(t)$\;
    }
    $\ell_{\max}\leftarrow \max_j \ell(j)$\;
    $\tilde{\alpha}_i(t)\leftarrow \ell_{\max}+\log\!\Big(\sum_j \exp(\ell(j)-\ell_{\max})\Big)$\;
  }
}

\BlankLine
\textbf{Backward pass:}\;
{\footnotesize (For causal inference, skip this part and set $\tilde{\beta}_i(t)\equiv 0$ for all $t$ and $i$ in the next block.)}\;
\For{$t\gets T-1$ \KwDownTo $0$}{
  \For{$i\gets 1$ \KwTo $2$}{
    \For{$j\gets 1$ \KwTo $2$}{
      $\ell(j)\leftarrow \tilde{\beta}_j(t+1)+\log A_{ij}+\log b_j(t{+}1)$\;
    }
    $\ell_{\max}\leftarrow \max_j \ell(j)$\;
    $\tilde{\beta}_i(t)\leftarrow \ell_{\max}+\log\!\Big(\sum_j \exp(\ell(j)-\ell_{\max})\Big)$\;
  }
}

\BlankLine
\textbf{Posterior probabilities:}\;
\For{$t\gets 0$ \KwTo $T$}{
  \For{$i\gets 1$ \KwTo $2$}{
    $\gamma(i)\leftarrow \tilde{\alpha}_i(t)+\tilde{\beta}_i(t)$\;
  }
  $\gamma_{\max}\leftarrow \max_i \gamma(i)$\;

  \For{$i\gets 1$ \KwTo $2$}{
    $\log p(S(t){=}S_i\mid y(0{:}T)) \leftarrow
    \gamma(i)-\gamma_{\max}-\log\!\Big(\sum_k \exp(\gamma(k)-\gamma_{\max})\Big)$\;
    $p(S(t){=}S_i\mid y(0{:}T))\leftarrow \exp\!\big(\log p(S(t){=}S_i\mid y(0{:}T))\big)$\;
  }
}

\Return $p(S(t)=S_i\mid y(0{:}T))$, $\forall t$\;
\end{algorithm}

\paragraph{Emission Model}
\label{emission}
To perform HMM inference, we specify the emission (observation) model
$p(y(t)\mid S(t))$, where $y(t)$ denotes the scalar NET-based observation at time $t$.
Specifically, the NET observation is obtained from the correlation-based metric after
applying the Fisher transformation
\begin{equation}
y(t) = \mathrm{atanh}\big(r(t)\big),
\label{eq:fisher_transform}
\end{equation}
where $r(t)$ denotes the raw Pearson correlation coefficient. After the Fisher
transformation, correlation coefficients are approximately normally distributed~\cite{fisher1915frequency,geirnaert2026modeling},
which motivates the use of Gaussian emission models.

Accordingly, we model the emission probability $b_i(t)$ required by the inference algorithm as:
\begin{equation}
p\big(y(t)\mid S(t)=S_i\big)
=
\mathcal{N}\!\big(y(t);\mu_i,\sigma^2\big),
\label{eq:emission_gauss_scalar}
\end{equation}
where $\mu_i$ is the state-dependent mean and $\sigma^2$ is a variance shared across both states.
The tied-variance assumption reduces the number of free parameters and improves numerical
robustness, and was shown to be realistic in sAAD \cite{lopez2025unsupervised}.

Under the Gaussian emission assumption in \eqref{eq:emission_gauss_scalar}, the emission parameters
can be estimated directly from the observed sequence in an unsupervised manner. Specifically, for
each participant, we use a multiscale two-component Gaussian mixture model (GMM) with shared variance
to estimate the emission parameters from observations computed at multiple window lengths
($5$, $10$, and $30$~s), following the multiscale fitting strategy in \cite{li2026multiscalegaussianmixturemodelinghmm}.
The key idea is to estimate common state-dependent means while accounting for the fact that the
variance of Fisher-transformed correlation estimates decreases with increasing window length.
Accordingly, the GMM estimates shared component means and a common variance scale, which is then
converted to the emission variance corresponding to the analysis window length used for HMM inference.
The GMM parameters are estimated using the expectation-maximization (EM) algorithm
\cite{dempster1977maximum}. The resulting component means and window-length-adjusted variance are
then used to define the HMM emission parameters $(\mu_1,\mu_2,\sigma^2)$.

\section{Experiment and Evaluation}
\label{sec_exp}

\subsection{Protocol Design}
\begin{figure}[h]
    \centering
    \includegraphics[width=1.2\linewidth]{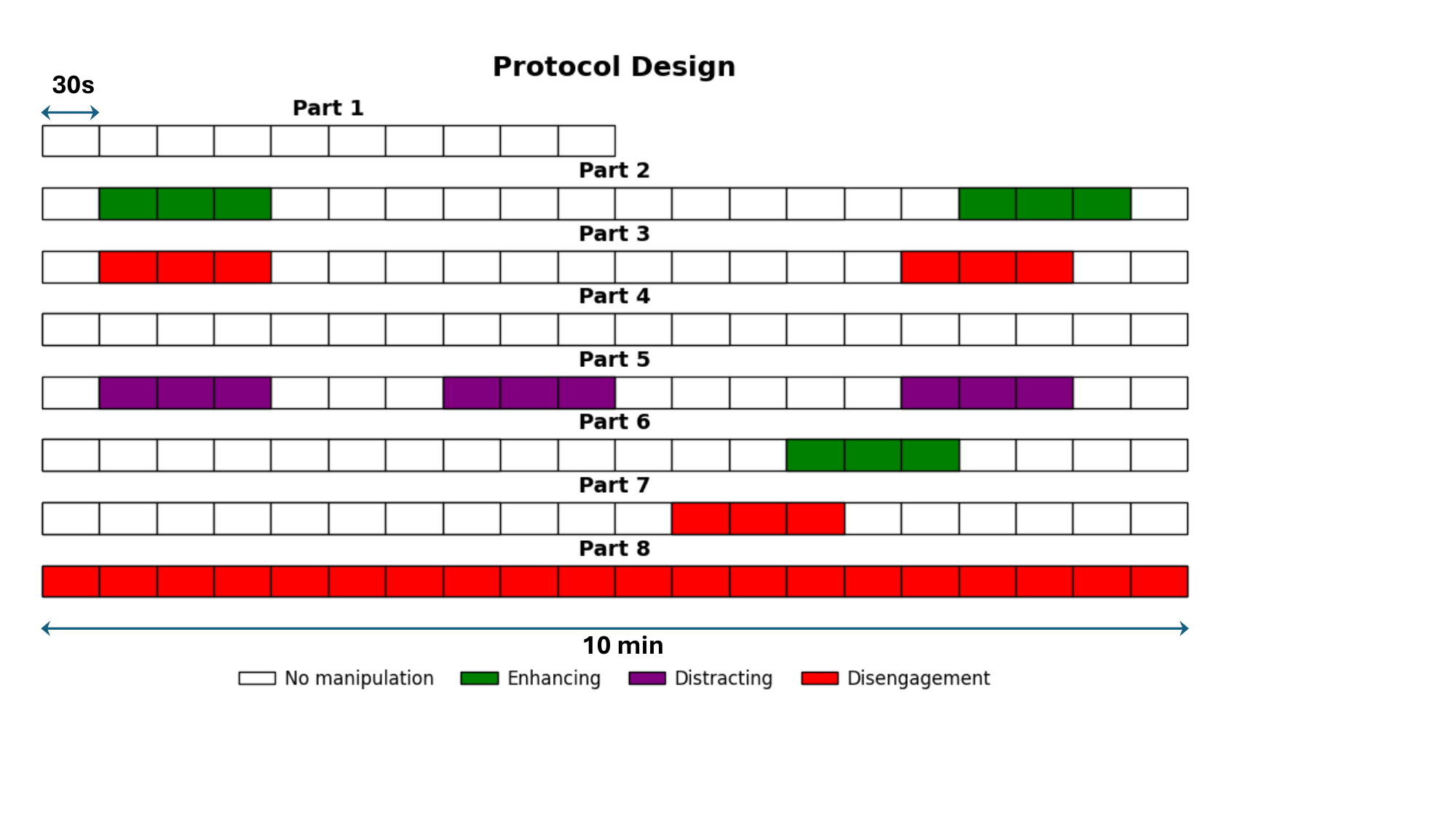}
    \caption{An overview of the experimental protocol \cite{vandenbroek_bellon_li_geirnaert_bertrand_desmedt_2025}. Each box corresponds to 30 seconds. White, red, green, and purple boxes indicate the no-manipulation, disengagement, attention-enhancing, and distracting conditions, respectively.}
    \label{fig:protocol}
\end{figure}

The experimental protocol is described in detail in \cite{vandenbroek_bellon_li_geirnaert_bertrand_desmedt_2025}. Briefly, EEG was recorded while 30 participants watched a 75-minute video lecture on ``neuromyths'', common but false beliefs about the brain. The video only shows the teacher in front of PowerPoint slides (mainly images with hardly any text) and the content was explained entirely through speech. The video was divided into eight parts of 10 minutes each (except the first part, which is only 5 minutes), separated by short breaks (see Fig. \ref{fig:protocol}).


The experiment included four distinct conditions to explore the dynamics of
auditory attention, as shown in Fig.~\ref{fig:protocol}:
\begin{description}[
    style=nextline,
    leftmargin=0pt,
    labelindent=0pt,
    topsep=0pt,
    partopsep=0pt,
    parsep=0pt,
    itemsep=0.4em
]
    \item[\textbf{No manipulation}]
    Participants were instructed to actively listen to the teacher without any
    additional prompts or interruptions.

    \item[\textbf{Disengagement}]
    Participants performed an interfering task to divert their attention away
    from the speaker. These tasks included solving calculation exercises,
    reading a text, or counting backwards in steps of 17 starting from 994.

    \item[\textbf{Distracting}]
    External distractors were introduced, including added background noise or
    a silent animal video displayed on the screen. However, the participants were still supposed to pay attention to the teacher's voice.

    \item[\textbf{Enhancing}]
    Visual cues (a red frame and an exclamation mark) were presented to indicate
    segments that would be included in the post-hoc comprehension questions,
    encouraging increased attentional engagement.
\end{description}

Unless stated otherwise, the analyses in this work primarily focus on the `no manipulation' and `disengagement' conditions (white and red boxes in Fig. \ref{fig:protocol}), which represent the clearest attentive (baseline) and inattentive listening state, respectively. The distracting and enhancing conditions are not removed to preserve a continuous recording, but they are ignored in the accuracy metrics.

\subsection{Data Acquisition and Preprocessing}
\label{data_acq}
EEG signals were recorded using a BioSemi Active Two system with 64 electrodes arranged according to the 10-20 system, at a sampling rate of 2048~Hz. Prior to analysis, the recordings were processed using a standard preprocessing pipeline.

To remove slow drifts and very low-frequency components, we first applied a first-order infinite impulse response (IIR) high-pass filter with a cutoff frequency of 0.01~Hz. The data were then downsampled to 128~Hz to reduce computational load while preserving the relevant EEG information.

The EEG signals were further filtered to the 0.1-4~Hz frequency band proposed in \cite{vanthornhout2019effect,EMBC}, which corresponds to typical word-rate dynamics around 2.5~Hz. This band was extracted using a fourth-order IIR bandpass filter, after which the signals were downsampled to 16~Hz for subsequent processing.

The speech envelope was adopted as the acoustic feature, motivated by previous findings showing phase-locked neural responses to speech envelope fluctuations \cite{geirnaert2021electroencephalography,lesenfants2020interplay,roebben2024you,vanthornhout2019effect,kubanek2013tracking,biesmans2016auditory,o2015attentional,EMBC}. Envelope extraction followed the procedure described in \cite{biesmans2016auditory}, which combines a gammatone filterbank with power-law compression to approximate cochlear frequency selectivity and nonlinear loudness perception.

For consistency with the EEG processing, the extracted envelope was filtered using the same frequency band (0.1-4~Hz by default) and downsampled to 16~Hz.

\subsection{Evaluation Metrics}

In this work, performance is evaluated based on the ability of the resulting
metrics (either the raw NET values or the HMM outputs) to discriminate between
the `no manipulation' attentive and `disengagement' inattentive conditions.
We assess performance along two complementary dimensions: \textit{steady-state discriminability}, which quantifies how well the two conditions can be discriminated from each other over the course of the session, and \textit{transition timing}, which quantifies how accurately the timing of attention-state changes is captured. The former is captured by the Area Under the Curve (AUC), while the latter is captured by the Switch Detection Time (SDT). Unless stated otherwise, AUC is used as the primary performance metric. SDT is reported only in analyses where transition timing is directly relevant.



\subsubsection{Area Under the Curve}

The Area Under the Receiver Operating Characteristic Curve (AUC) provides a
ranking-based measure of separability. It can be interpreted as the probability
that a randomly selected attention metric value from the baseline (attentive) condition is larger than
a randomly selected attention metric value from the disengagement (inattentive) condition.

Formally, given samples $y_1$ from the baseline condition and $y_2$ from the
disengagement condition, we estimate the AUC using its rank-based (Mann-Whitney)
formulation \cite{mann1947test,mason2002areas}:
\begin{equation}
\mathrm{AUC}
=
\frac{
\#(y_1 > y_2)
+
0.5\,\#(y_1 = y_2)
}{
\#(y_1)\,\#(y_2)
},
\label{eq:auc}
\end{equation}
where $\#(\cdot)$ denotes the number of occurrences meeting the specified condition. A higher AUC value indicates stronger discrimination between the
two experimental conditions and, consequently, improved aAAD performance.

\subsubsection{Switch Detection Time\protect\footnotemark}
\footnotetext[2]{The terminology SDT was introduced in
\cite{heintz2025postprocessingeegbasedauditoryattention} in the context of
sAAD, where attention is literally ``switched'' between different sound
sources. In our aAAD context, ``switch'' is a misnomer, yet we retain this
terminology for consistency with that work.}
The SDT quantifies how accurately the estimated attention trajectory captures
the timing of protocol-defined transitions between the `no manipulation' and
`disengagement' conditions. In this work, we use the optimal split criterion
employed in the evaluation protocol of the ICASSP 2027 EEG-sAAD Grand Challenge
\cite{icassp2027eegsaad}. Whereas the challenge considers causal detection,
we adapt the criterion to our non-causal setting by using a symmetric search
interval around each transition. The idea is to estimate the transition time by matching the predicted
state sequence to templates with different transition times. Each
template assigns the old state before a candidate time and the new
state from that time onward. We select the best-matching valid
template and take its transition time as the estimated transition
time. SDT is then the absolute difference between this estimate
and the ground-truth transition time.

Specifically, let $t_{k}$ denote the time of ground-truth transition $k$.
Each analysis window is timestamped by its end time. Since different
window lengths produce predictions at different time intervals, we
linearly interpolate the continuous HMM posterior trajectories onto
a common 1-s grid for SDT evaluation. The interpolated trajectories are then thresholded at 0.5 to obtain
binary state predictions.

We consider candidate split points $\xi$ on this grid within
$[t_{k}-45\text{~s},t_{k}+45\text{~s}]$. Let $\mathcal{T}_k$ denote the set of time points within this interval, $\hat{S}(t)$ the predicted
binary state, and $S_k^{-}$ and $S_k^{+}$ the true states before and
after transition $k$, respectively. The agreement score is
\begin{equation}
A_k(\xi)
=
\sum_{\substack{t\in\mathcal{T}_k\\t<\xi}}
\mathbf{1}\{\hat{S}(t)=S_k^{-}\}
+
\sum_{\substack{t\in\mathcal{T}_k\\t\geq\xi}}
\mathbf{1}\{\hat{S}(t)=S_k^{+}\},
\end{equation}
where $\mathbf{1}\{\cdot\}$ is the indicator function. We define the detected time for transition $t_k$ as the value for $\xi$ that has the highest agreement score. A candidate is valid only if the five consecutive predictions
starting at $\xi$ all indicate $S_k^{+}$ and lie within the search
interval. We select the valid candidate with the highest agreement
score as $\xi_k^*$. If several candidates have the same highest
score, we select the earliest. The SDT for transition $k$ is
\begin{equation}
\label{SDT}
\mathrm{SDT}_k = |\xi_k^*-t_{k}|.
\end{equation}
If no valid candidate exists, the transition is recorded as a miss
and excluded from SDT (SDT is only meaningful on detected transitions). Since we also consider non-causal systems, the detected transition can happen before the ground-truth transition, which is why we take the absolute value in \eqref{SDT} to penalize any alignment error (positive or negative). We remark here that such missed transitions are still (heavily) accounted for
in the AUC evaluation and thus still influence the overall performance.

We report SDT for individuals and for the group. For individuals,
we compute the median SDT over detected transitions for each participant
and then average these medians across participants. For the group,
we first average the continuous HMM trajectories across participants and
then compute the mean SDT over detected transitions. In paired
comparisons between two methods, only transitions detected by both methods are included.
For both individual and group SDT, 95\% confidence intervals
are obtained by bootstrapping participants 1000 times. 


\subsection{Default Experiment Settings}
\label{sec:default_settings}

Unless stated otherwise, the following default settings are used throughout the experiments. Non-causal (forward-backward) HMM inference is used, and the emission parameters are estimated per participant via the multiscale two-component GMM \cite{li2026multiscalegaussianmixturemodelinghmm} described in Section~\ref{emission} (`Emission model'), using fitting window lengths of $5$, $10$, and $30$~s. The default transition probability is $P_{\mathrm{trans}}=0.01$, which corresponds to an expected state persistence of $1/P_{\mathrm{trans}}=100$ time steps. Since one time step corresponds to an analysis window of $N$ samples (corresponding to $W=N/f_s$ seconds)
, the expected state persistence in physical time is $W/P_{\mathrm{trans}}$ seconds. For a 1-s window, this translates to an expected dwell time of 100~s. When the window length is varied (Section~\ref{sec_tradeoff}), $P_{\mathrm{trans}}$ is scaled to maintain a constant expected dwell time in physical time, i.e.,
\[
\frac{W}{P_{\mathrm{trans}}}=100 \text{~s},
\]
e.g., $P_{\mathrm{trans}}=0.3$ for $W=30$~s. Since we use $P_{\mathrm{trans}}=0.01$ as the default setting, the corresponding default window length is $W=1$~s. The influence of the window length and $P_{\mathrm{trans}}$ on AUC and SDT is investigated in Sections~\ref{sec_tradeoff} and~\ref{sec:robust_irregular}, respectively.

\section{Results and discussion}
\label{sec:results}

\subsection{Added value of HMM}
\label{subsec:benefits_hmm}

\begin{figure}[h!]
    \centering
    \includegraphics[width=\linewidth]{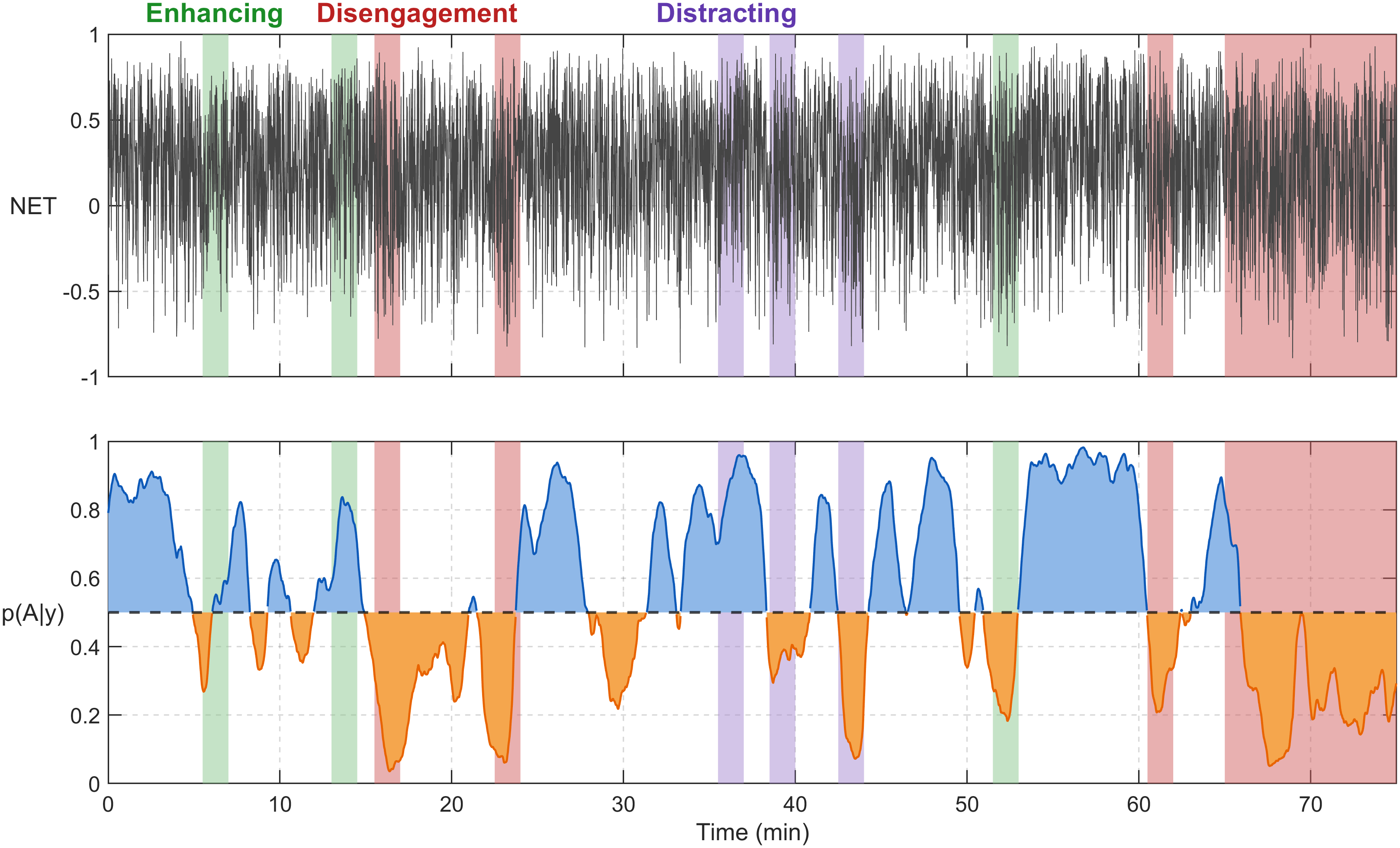}

    {\footnotesize (a) 75th percentile (participant~13)\par}
    \vspace{0.5em}

    \includegraphics[width=\linewidth]{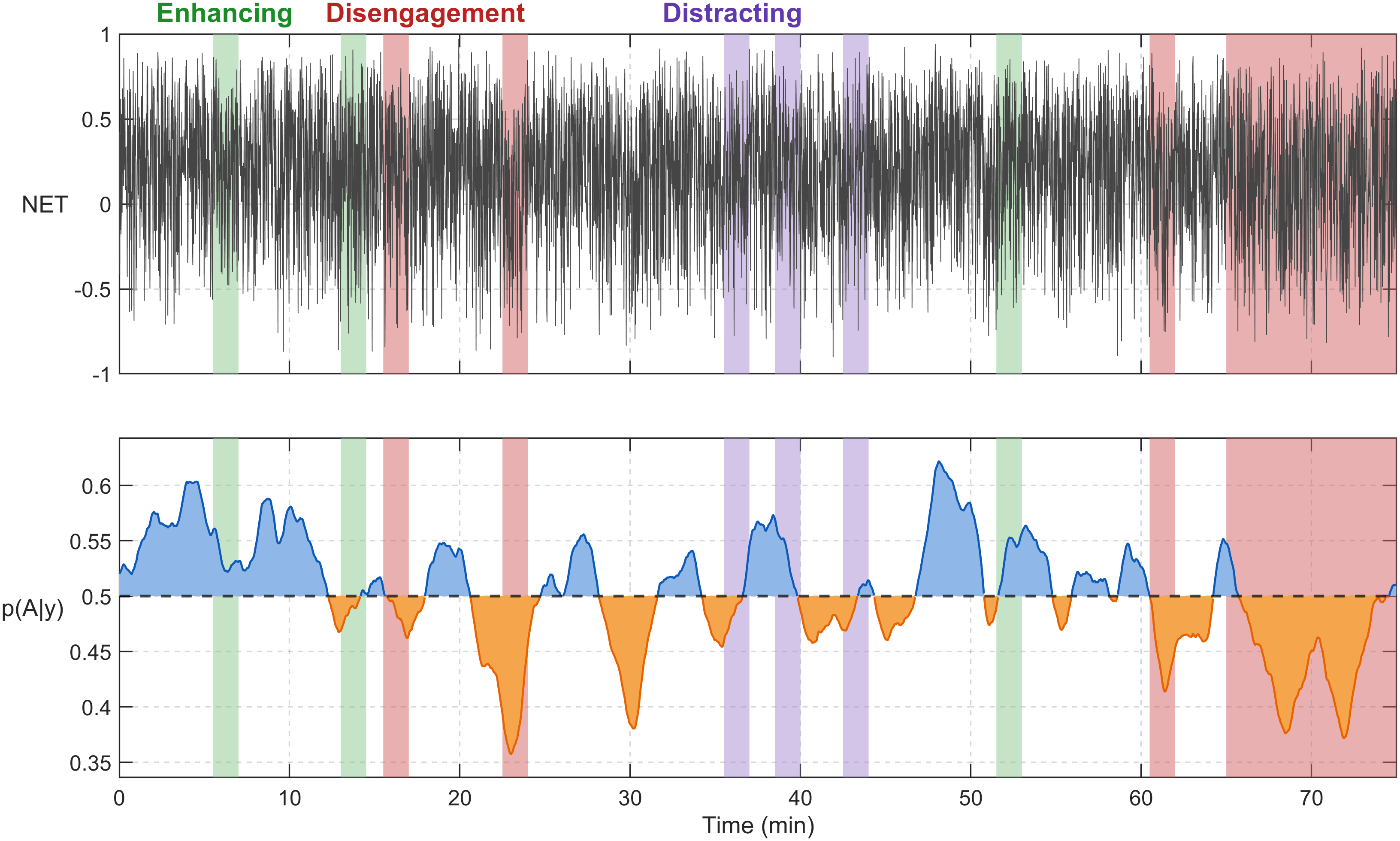}

    {\footnotesize (b) 50th percentile (participant~27)\par}
    \vspace{0.5em}

    \includegraphics[width=\linewidth]{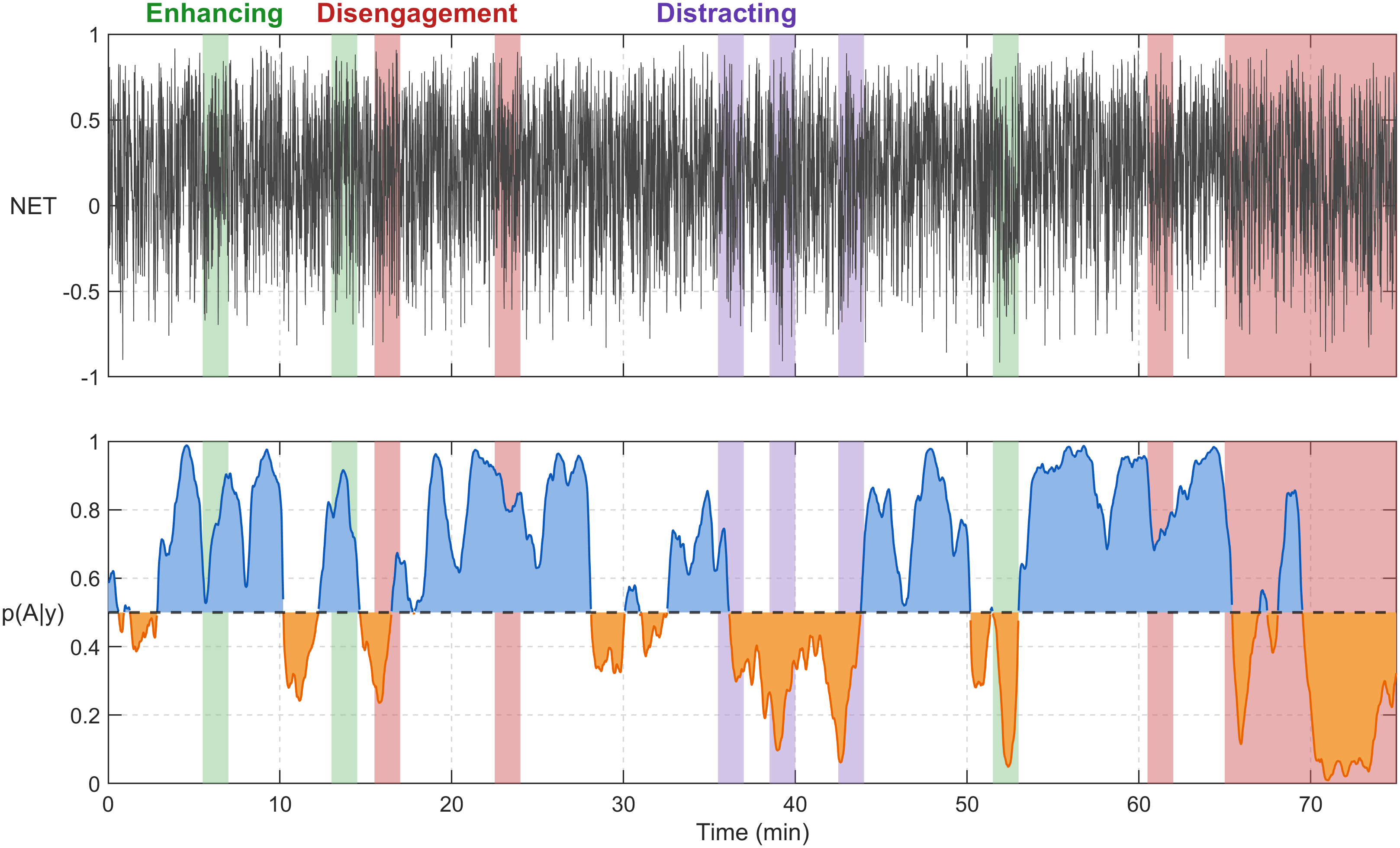}

    {\footnotesize (c) 25th percentile (participant~4)\par}

    \caption{Comparison of raw NET traces (upper panels) and non-causal HMM posteriors (lower panels) over 75-minute sessions (1-s windows) for participants at the 75th, 50th, and 25th AUC percentiles. The dashed line at 0.5 serves as a reference threshold. Blue and orange shading denote periods where the posterior probability of being attentive is larger than 0.5 or smaller than 0.5, respectively. Higher-performing participants yield HMM posteriors that more accurately track the expected attentional transition, i.e., fewer false positives (blue scores) within the disengagement conditions (red background). Relative to the protocol labels, orange scores within baseline segments (white background) are false negatives, although they may also reflect genuine but unobserved lapses of attention.}
    \label{fig:traces}
\end{figure}

To illustrate the added value of HMM post-processing, we compare the Fisher-transformed raw NET trace with the corresponding HMM-processed trace obtained via non-causal inference, which provides an off-line estimate of the latent attentional state. Fig.~\ref{fig:traces} presents three example 75-minute sessions (NET scores computed over 1-s windows) for participants at the 75th (participant~13), 50th (participant~27), and 25th (participant~4) percentiles of AUC performance, respectively. For each participant, the upper panel displays the raw NET and the lower panel shows the HMM posterior probability of the attentive state, $p(S(t)=A \mid y(0:T))$. The dashed horizontal line at 0.5 serves as a reference threshold (which could serve as a decision boundary if binary classification were desired, although the present work treats the posterior as a continuous measure of attentional engagement). Blue and orange shading indicate periods where the posterior lies above and below this reference, respectively, providing a visual summary of how attentional engagement evolves over time.
Across subjects at different performance quantiles, the raw NET shows strong temporal variability, making it difficult to identify a clear structure over time. In contrast, the HMM output provides a smoother and more interpretable trajectory. The HMM posterior generally remains high during `no manipulation' periods (white background) and decreases during `disengagement' periods (red background). The former can be viewed as attentive periods and the latter as inattentive periods, although it is important to note that these should not be treated as a fully reliable ground truth. This is particularly true for the `no manipulation' segments, because participants could still be inattentive during these baseline segments.

\begin{figure}[t]
    \centering
    \includegraphics[width=1\linewidth]{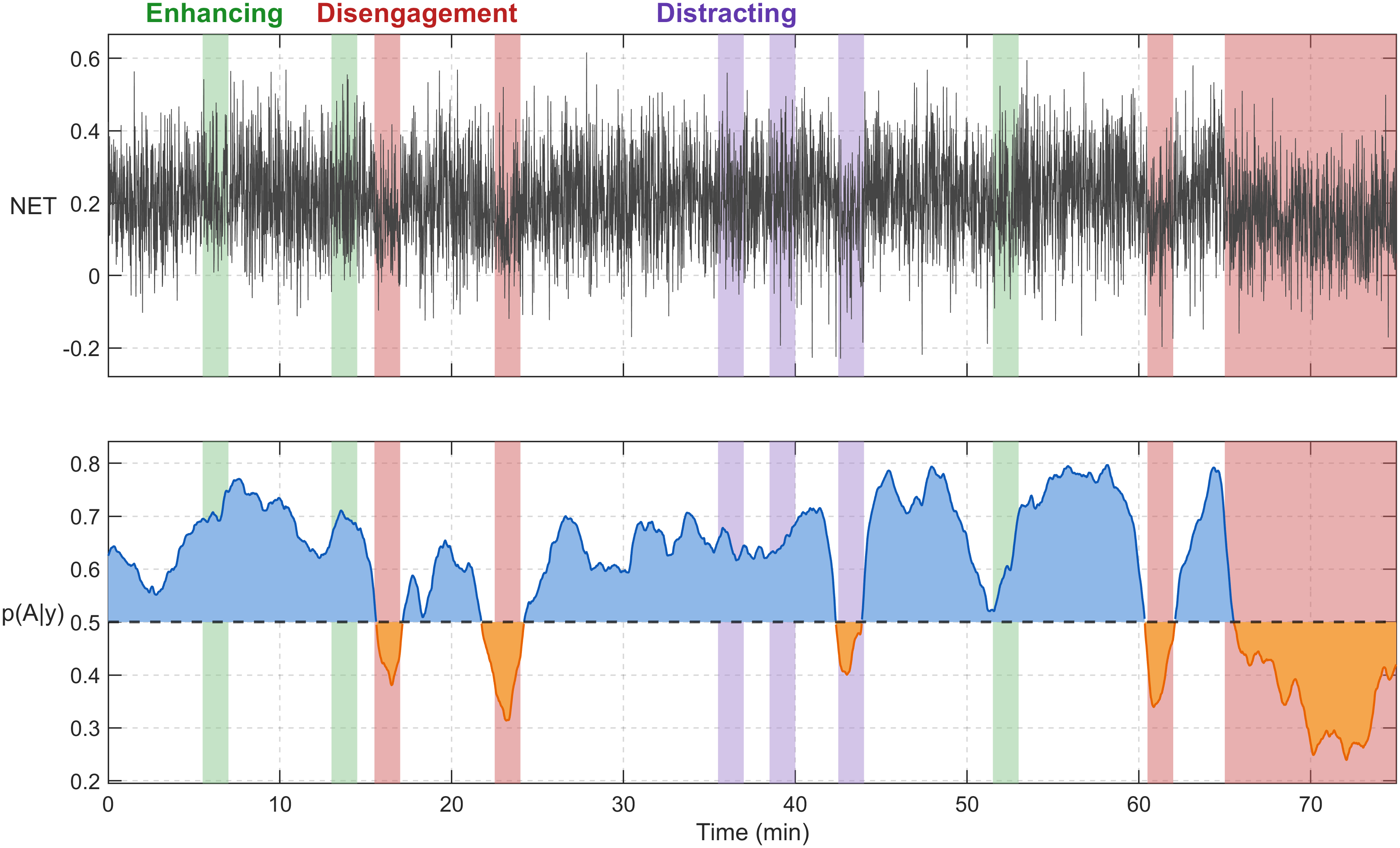}
    \caption{Group-averaged NET trace and non-causal HMM posterior over the full session (1-s windows). Even after averaging across participants, the raw NET remains relatively noisy, whereas the HMM posterior reveals a clearer temporal structure aligned with the manipulation schedule.}
    \label{fig:trace_compare}
\end{figure}

To assess whether this behavior generalizes beyond a single participant, Fig.~\ref{fig:trace_compare} shows the traces averaged across all participants. Although averaging reduces some of the variability in the raw NET, the group-level NET trace remains relatively noisy and does not clearly reflect the manipulation schedule. In contrast, the averaged HMM output exhibits a clearer trend and more consistent drops during distracting and disengagement periods, indicating that the temporal inference step extracts a more robust group-level attention profile.

\begin{figure}[t]
    \centering
    \includegraphics[width=0.8\linewidth]{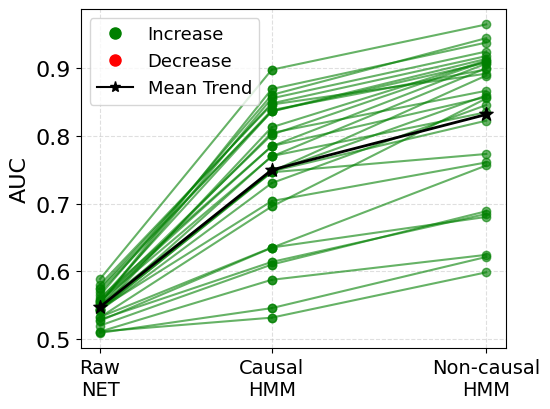}
    \caption{AUC improvement per participant. Each point represents one participant. All participants benefit from HMM post-processing ($P_{\mathrm{trans}}=0.01$) for a 1-s analysis window.}
    \label{fig:AUCbenefit}
\end{figure}

Beyond visual inspection, we quantify the benefit at the individual level using AUC. Fig. \ref{fig:AUCbenefit} reports per-participant performance for the raw NET and for the HMM outputs. Each marker corresponds to one participant. Both causal and non-causal HMM inference outperform the raw NET baseline across participants, and the non-causal configuration consistently yields higher performance, as expected because it can leverage future observations when estimating the current state.

We focus this comparison on AUC rather than SDT, because the goal here is to quantify the gain in discriminability introduced by HMM post-processing. SDT is not reported for raw NET because the evaluation requires binary state predictions, whereas raw NET is a noisy evidence sequence without a natural decision threshold.

It is important to clarify that the HMM does not replace NET as an alternative decoding method. NET provides the window-wise neural evidence, while the HMM serves as a temporal post-processing stage that integrates this evidence over time by explicitly modeling state persistence. The observed performance gains should therefore be interpreted as the benefit of structured temporal integration of the same NET sequence, rather than as a comparison between independent decoding approaches.

Furthermore, the `causal' configuration considered here refers only to the temporal inference stage of the HMM. The preceding steps in the pipeline, including EEG and speech preprocessing, neural decoder training, and emission distribution estimation, remain non-causal. Consequently, the reported causal performance represents an upper bound relative to a fully real-time implementation. In a strictly causal end-to-end setting, where all components operate without access to future data, performance could be lower.

\subsection{Effect of Window Length on AUC and SDT}
\label{sec_tradeoff}

\begin{figure}[!t]
    \centering
    \includegraphics[width=0.8\linewidth]{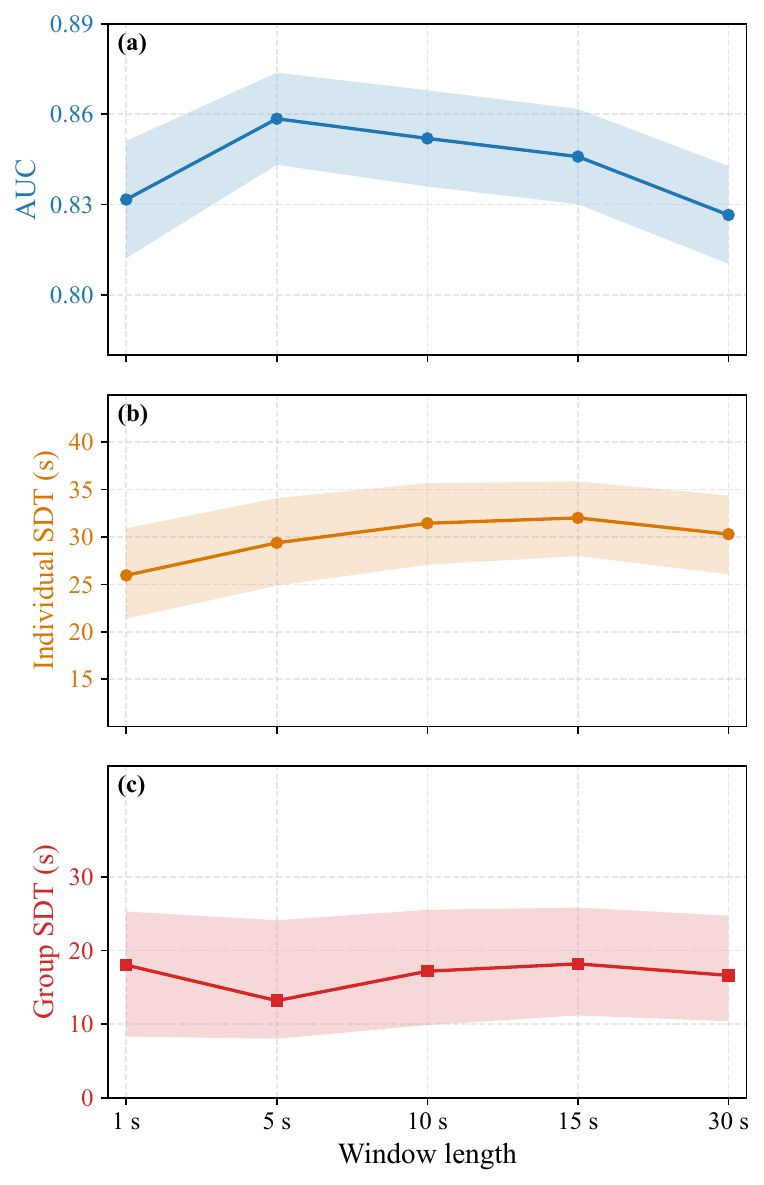}
    \caption{Effect of analysis window length on AUC and SDT.
    (a) Mean AUC across participants, with shading indicating $\pm1$
    standard error of the mean (SEM).
    (b) Individual SDT, computed by taking the median over detected
    switches for each participant and then averaging across participants.
    (c) Group-level SDT computed from the group-averaged HMM posterior
    trajectory by averaging over detected switches. Shaded regions in
    (b) and (c) indicate 95\% confidence intervals obtained by bootstrapping
    participants.}
    \label{fig:tradeoff}
\end{figure}

In this subsection, we investigate how the analysis window length affects both AUC and SDT for the non-causal HMM output. We consider window lengths of 1, 5, 10, 15, and 30~s. For each window length, AUC is evaluated across participants, whereas SDT is evaluated for individuals and for the group as described in Section~\ref{sec_exp}. Fig.~\ref{fig:tradeoff} shows the results. Mean AUC increases from 0.832 at 1~s to its highest value of 0.858 at 5~s, then gradually decreases to 0.827 at 30~s. Individual SDT ranges from 25.95 to 32.02~s, and the proportion of missed transitions across participants ranges from 29.0\% to 31.0\%. The 1-s window yields both the lowest individual SDT and the lowest miss rate. Group SDT ranges from 13.17 to 18.17~s, with the lowest value at 5~s. The numbers of missed group transitions for window lengths of 1, 5, 10, 15, and 30~s are zero, one, one, one, and two out of seven, respectively. The 1-s window detects all seven group transitions with a group SDT of 18.00~s. Together with its mean AUC of 0.832 and its performance for individuals, these results support the use of 1~s as the default window length.

\subsection{Comparison with MA Smoothing}
\label{sec:robust_irregular}

In this subsection, we compare the HMM with a non-causal moving-average (MA)
baseline, which represents a basic fixed-window smoothing approach. This
analysis is performed for the default 1-s analysis window, where the NET
sequence has the highest temporal resolution.

For a Fisher-transformed NET sequence $y(t)$ and an even smoothing parameter
$\tau$, the MA output is defined as
\[
\hat{y}_{\mathrm{MA}}(t)
=
\frac{1}{\tau+1}
\sum_{k=-\tau/2}^{\tau/2} y(t+k).
\]
This corresponds to a symmetric smoothing window containing $\tau/2$ past
samples, the current sample, and $\tau/2$ future samples, resulting in a total
window length of $\tau+1$ samples. The use
of both past and future samples matches the non-causal setting of the
forward-backward HMM inference used in the main analysis.

To use the same smoothing parameter for MA and HMM, we set
\[
\tau = \frac{1}{P_{\mathrm{trans}}},
\]
where $1/P_{\mathrm{trans}}$ is the expected state dwell time of the HMM in
discrete time steps. Under this parameterization, $\tau$ controls the smoothing
scale for MA through the averaging-window length, and for the HMM through the
expected state persistence. These mechanisms are not equivalent, but using the
same parameter allows us to compare how both methods behave as the smoothing
scale is varied.

We evaluate both methods over a range of smoothing strengths
($\tau=10$-$160$~s). The default value used elsewhere in the paper,
$\tau=100$~s, corresponds to $P_{\mathrm{trans}}=0.01$ and was chosen a priori
to reflect sustained attention-state persistence; it was not optimized using
the evaluation labels. AUC is computed at the participant level from the
continuous HMM posterior or MA-smoothed NET score. SDT is computed at both the
individual and group levels using the procedure described in
Section~\ref{sec_exp}. The HMM posterior is thresholded at 0.5. Because the MA
output is not a posterior probability, its threshold is determined in an
unsupervised manner by fitting a two-component GMM and using the midpoint
between the two component means. The GMM is fitted separately to each
participant's MA trajectory for the individual-level analysis and to the
group-averaged MA trajectory for the group-level analysis.

\begin{figure}[!t]
    \centering
    \includegraphics[width=0.8\linewidth]
    {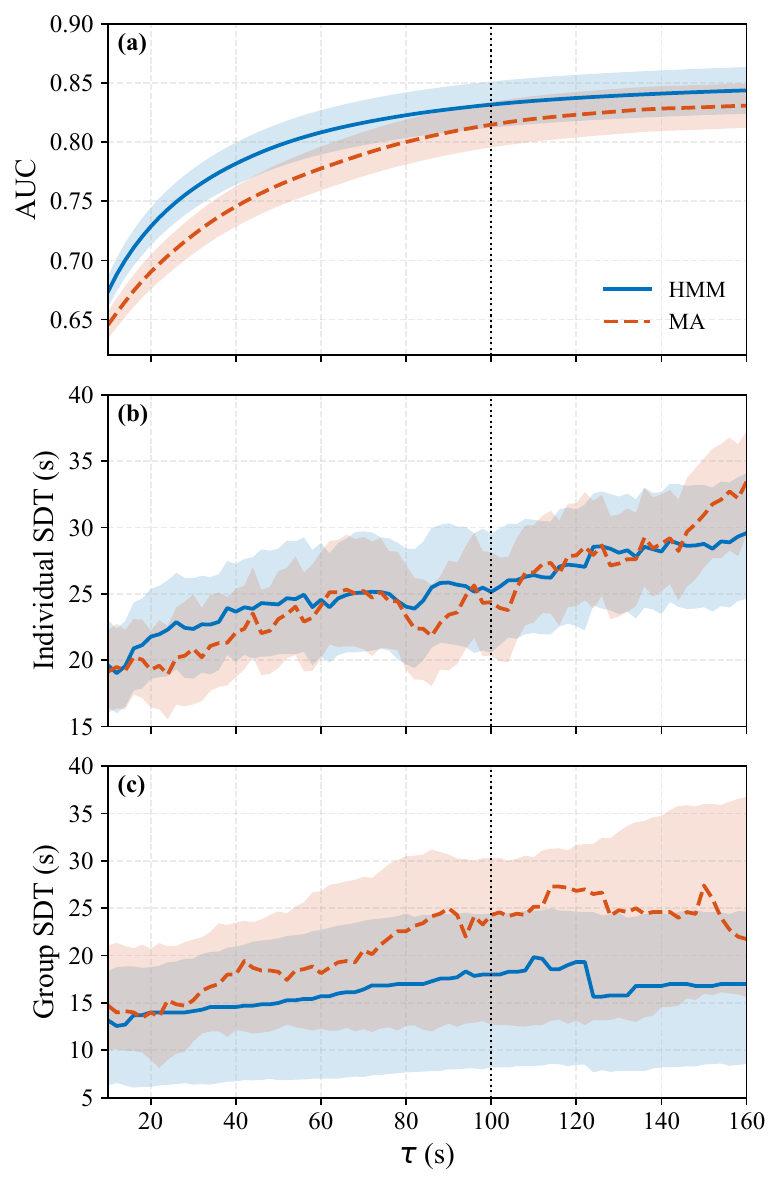}
    \caption{Comparison between HMM inference and non-causal MA smoothing
    for the 1-s analysis window. (a) AUC. (b) Individual SDT.
    (c) Group-level SDT. AUC and SDT are shown as functions of the
    smoothing scale $\tau$. For the HMM, $\tau=1/P_{\mathrm{trans}}$;
    for MA, $\tau$ defines the symmetric averaging scale. The vertical
    dotted line marks the default value $\tau=100$~s. AUC shading
    indicates $\pm1$ SEM across participants. Individual SDT is
    computed by taking the median over switches detected by both methods
    for each participant and then averaging across participants.
    Group-level SDT is computed from the group-averaged trajectories by
    averaging over switches detected by both methods. SDT shading
    indicates 95\% participant-bootstrap confidence intervals.}
    \label{fig:hmm_vs_ma_tau}
\end{figure}

Fig.~\ref{fig:hmm_vs_ma_tau} summarizes the comparison. Across the examined
range of smoothing strengths, the HMM consistently achieves higher AUC than
the MA baseline, indicating that its advantage is not merely due to generic
temporal smoothing. Individual SDT is similar for the two methods, with no
consistent advantage for either. At the default $\tau=100$~s, individual SDT over jointly detected
transitions is 25.15~s for the HMM and 24.37~s for MA. The HMM and MA miss 61 out of
210 transitions (29.0\%) and 51 out of 210 (24.3\%), respectively. Across
the full range of smoothing scales, the corresponding miss rates increase
from 2.4\% to 33.3\% for the HMM and from 1.4\% to 34.8\% for MA.

For the group, the HMM yields lower SDT than MA over most of the examined
range. At the default $\tau=100$~s, group SDT is 18.00~s for the HMM and
24.29~s for MA, and both methods detect all seven transitions. The HMM
detects all seven group transitions across the full range of smoothing
scales, whereas MA misses up to three. These results show that the HMM
achieves higher AUC and generally lower group SDT than fixed-window
averaging, while individual SDT remains comparable.

\subsection{Why NET is Particularly Suitable as Evidence for aAAD}
\label{sec:feature_comparison}

The proposed HMM framework is presented using NET as the observation feature throughout this paper. Here, we justify this choice by comparing NET with two commonly used EEG-based attention markers, namely alpha power \cite{KLIMESCH199873} and spectral entropy \cite{lesenfants2020interplay}, using the same HMM framework. A key distinction lies in the type of attention captured by each of these features. NET directly reflects neural tracking of the attended speech stimulus and therefore provides a clear and consistent measure of auditory attention to a specific target speech signal. In contrast, alpha power and entropy are more general neural markers that have been linked to overall attention in a broader sense (e.g., arousal or cognitive effort), but that are not specifically tied to attention toward a particular auditory source.

Specifically, alpha power is defined as the mean band power of the EEG signal in the 8-12~Hz range \cite{KLIMESCH199873}, averaged across all electrodes. It is commonly used as a marker of attentional state \cite{KLIMESCH199873, benedek2014alpha}, although the direction of modulation can vary across individuals and paradigms \cite{roebben2024you}. Spectral entropy reflects the regularity of the EEG signal and is commonly computed from the power spectrum over a broad frequency range (0.5-32~Hz in \cite{lesenfants2020interplay,roebben2024you}). The power spectrum, estimated using Welch's method, is first normalized to form a probability distribution. Spectral entropy is then computed as 
\cite{lesenfants2020interplay,roebben2024you}
\begin{equation}
SE = -\sum_k p_k \log_2 p_k,
\end{equation}
where $p_k$ denotes the normalized power at frequency bin $k$. Higher spectral entropy indicates a more uniform spectral distribution, whereas lower values reflect a stronger concentration of power in specific frequency bands. Both alpha power and spectral entropy are computed per analysis window using the same window lengths as NET and averaged across all electrodes, but notably they do not use the speech signal in their computation. As a result, their relationship with auditory attention is ambiguous, and the direction of modulation (relative to the other conditions or tasks) is not well defined, which makes it hard to assign class labels in an unsupervised setting. In the proposed fully unsupervised GMM-based emission estimation for NET, the component with the larger mean is assigned to the attentive state. This assignment is naturally justified, as higher neural tracking scores consistently indicate stronger auditory attention. However, the same assignment is not necessarily valid for alpha power and entropy, for which the direction of modulation may even vary across participants. To provide a fair comparison, we determine the correct direction for each feature using the ground-truth labels, i.e., based on the average ordering between attentive and inattentive segments across all participants. Specifically, higher alpha power and lower entropy are associated with attentive state, and this is used consistently in the HMM model. However, we note that this is based on the average across participants, and this direction is not always fully consistent across participants.

We compare the three features using non-causal HMM inference with the
same multiscale emission estimation procedure and transition probabilities
as in the preceding analysis. We consider window lengths of 1, 5, 10,
15, and 30~s. AUC and SDT are computed as described in
Section~\ref{sec_exp}. The figure shows SDT over transitions detected
by each feature.

\begin{figure}[!t]
    \centering
    \includegraphics[width=0.8\linewidth]{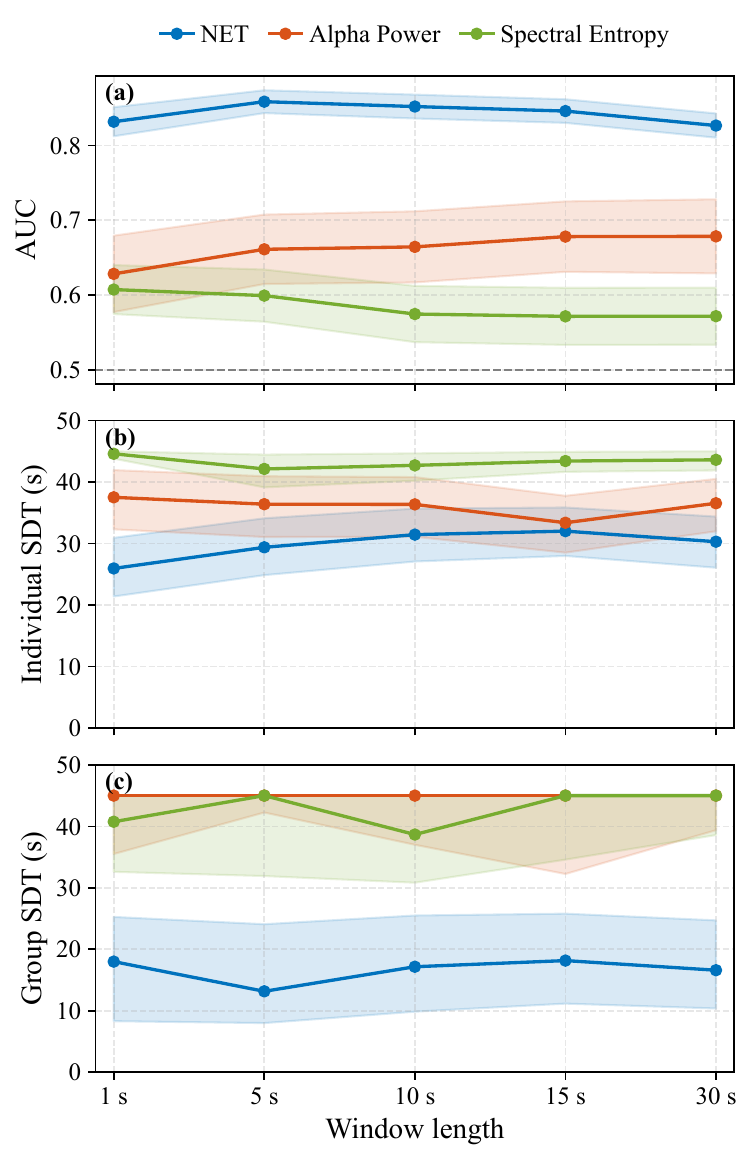}
    \caption{Comparison of HMM decoding performance using NET, alpha power,
    and spectral entropy across analysis window lengths. (a) Mean AUC,
    with shading indicating $\pm1$ SEM across 30 participants; the dashed
    line marks chance level (AUC\,=\,0.5). (b) Individual SDT.
    (c) Group SDT. Each SDT curve uses transitions detected by the
    corresponding feature. SDT shading indicates 95\% confidence
    intervals obtained by bootstrapping participants.}
    \label{fig:feature_comparison}
\end{figure}

Fig.~\ref{fig:feature_comparison} summarizes the feature comparison.
NET achieves the highest mean AUC across all tested window lengths
(Fig.~\ref{fig:feature_comparison}a), whereas alpha power and spectral
entropy show weaker discriminability. At the default 1-s setting,
the mean AUCs are 0.832, 0.628, and 0.607, respectively. NET also yields
the lowest individual and group SDT across all window lengths
(Fig.~\ref{fig:feature_comparison}b and c). At 1~s, individual SDT is
25.95~s for NET, 37.52~s for alpha power, and 44.60~s for spectral
entropy; the corresponding group SDTs are 18.00~s, 45.00~s (the
optimal split falls at the left search boundary because the mean
posterior stays below 0.5), and 40.75~s. The individual miss rates
are 29.0\%, 30.5\%, and 28.6\%, respectively, while the numbers of
missed group transitions are zero, three, and three out of seven.  These results support the use of
NET as the observation feature in the proposed unsupervised
HMM framework: unlike alpha power and spectral entropy,
NET is stimulus-specific and has a well-defined direction with
respect to auditory attention to the target speech.

\FloatBarrier

\section{Conclusion}
\label{sec_conc}
We have presented a practical, modular pipeline for unsupervised absolute auditory attention decoding (aAAD) in single-speaker listening scenarios. The proposed pipeline extracts a windowed neural envelope tracking (NET) sequence as stimulus-specific evidence and applies an HMM for probabilistic temporal inference, enabling continuous attention tracking without relying on explicit attention labels. Across all participants, HMM post-processing consistently improves the separability between `no manipulation' (attentive) and `disengagement' (inattentive) periods compared to the raw NET trace, yielding a more interpretable attention trajectory. We further investigate the effect of analysis window length on AUC and SDT. Compared with MA smoothing, the HMM achieves higher discriminability across the examined smoothing scales. Although the individual SDTs of the two methods are comparable, the HMM provides lower group-level SDT over most smoothing scales after participant averaging. Finally, we show that NET outperforms alpha power and spectral entropy in terms of AUC within the proposed framework, supporting its use as the observation feature for HMM-based attention tracking.



\section*{Data Availability Statement}
The data supporting the findings of this study are not publicly available and are not available for external sharing.

\section*{Ethics Statement}
The experimental protocol was approved by the Social and Societal Ethics Committee (SMEC) of KU Leuven (approval number: G-2023-7126). All participants provided written informed consent prior to participation. The study was conducted in accordance with the relevant ethical guidelines and regulations of the SMEC.

\section*{Conflict of Interest}
The authors declare no competing interests.

\bibliographystyle{IEEEtran}
\bibliography{ref}

@article{geirnaert2021electroencephalography,
  title={Electroencephalography-based auditory attention decoding: Toward neurosteered hearing devices},
  author={Geirnaert, Simon and Vandecappelle, Servaas and Alickovic, Emina and De Cheveigne, Alain and Lalor, Edmund and Meyer, Bernd T and Miran, Sina and Francart, Tom and Bertrand, Alexander},
  journal={IEEE Signal Processing Magazine},
  volume={38},
  number={4},
  pages={89--102},
  year={2021},
  publisher={IEEE}
}

@article{kubanek2013tracking,
  title={The tracking of speech envelope in the human cortex},
  author={Kubanek, Jan and Brunner, Peter and Gunduz, Aysegul and Poeppel, David and Schalk, Gerwin},
  journal={PloS one},
  volume={8},
  number={1},
  pages={e53398},
  year={2013},
  publisher={Public Library of Science San Francisco, USA}
}

@article{roebben2024you,
  title={‘{A}re you even listening?’-{E}{E}{G}-based decoding of absolute auditory attention to natural speech},
  author={Roebben, Arnout and Heintz, Nicolas and Geirnaert, Simon and Francart, Tom and Bertrand, Alexander},
  journal={Journal of Neural Engineering},
  volume={21},
  number={3},
  pages={036046},
  year={2024},
  publisher={IOP Publishing}
}

@article{KLIMESCH199873,
  title={Induced alpha band power changes in the human {EEG} and attention},
  author={Klimesch, Wolfgang and Doppelmayr, Michael and Russegger, Harald and Pachinger, Thomas and Schwaiger, Jens},
  journal={Neuroscience letters},
  volume={244},
  number={2},
  pages={73--76},
  year={1998},
  publisher={Elsevier}
}

@article{lesenfants2020interplay,
  title={The interplay of top-down focal attention and the cortical tracking of speech},
  author={Lesenfants, Damien and Francart, Tom},
  journal={Scientific Reports},
  volume={10},
  number={1},
  pages={6922},
  year={2020},
  publisher={Nature Publishing Group UK London}
}

@article{poulsen2017eeg,
  title={{EEG} in the classroom: Synchronised neural recordings during video presentation},
  author={Poulsen, Andreas Trier and Kamronn, Simon and Dmochowski, Jacek and Parra, Lucas C and Hansen, Lars Kai},
  journal={Scientific reports},
  volume={7},
  number={1},
  pages={43916},
  year={2017},
  publisher={Nature Publishing Group UK London}
}

@article{vanthornhout2019effect,
  title={Effect of task and attention on neural tracking of speech},
  author={Vanthornhout, Jonas and Decruy, Lien and Francart, Tom},
  journal={Frontiers in neuroscience},
  volume={13},
  pages={977},
  year={2019},
  publisher={Frontiers Media SA}
}

@article{etard2019decoding,
  title={Decoding of selective attention to continuous speech from the human auditory brainstem response},
  author={Etard, Octave and Kegler, Mikolaj and Braiman, Chananel and Forte, Antonio Elia and Reichenbach, Tobias},
  journal={NeuroImage},
  volume={200},
  pages={1--11},
  year={2019},
  publisher={Elsevier}
}

@article{biesmans2016auditory,
  title={Auditory-inspired speech envelope extraction methods for improved {EEG}-based auditory attention detection in a cocktail party scenario},
  author={Biesmans, Wouter and Das, Neetha and Francart, Tom and Bertrand, Alexander},
  journal={IEEE Transactions on Neural Systems and Rehabilitation Engineering},
  volume={25},
  number={5},
  pages={402--412},
  year={2016},
  publisher={IEEE}
}

@article{o2015attentional,
  title={Attentional selection in a cocktail party environment can be decoded from single-trial {EEG}},
  author={O'Sullivan, James A and Power, Alan J and Mesgarani, Nima and Rajaram, Siddharth and Foxe, John J and Shinn-Cunningham, Barbara G and Slaney, Malcolm and Shamma, Shihab A and Lalor, Edmund C},
  journal={Cerebral cortex},
  volume={25},
  number={7},
  pages={1697--1706},
  year={2015},
  publisher={Oxford University Press}
}

@misc{vandenbroek_bellon_li_geirnaert_bertrand_desmedt_2025,
  title={Tracking individual auditory attention during learning via {EEG}-based neural envelope tracking},
  author={Van den Broek, Davina and Bellon, Elien and Li, Tianyi and Geirnaert, Simon and Bertrand, Alexander and De Smedt, Bert},
  year={2025},
  howpublished={PsyArXiv preprint}
}

@INPROCEEDINGS{EMBC,
  author={Li, Tianyi and Geirnaert, Simon and Van Den Broek, Davina and Bellon, Elien and De Smedt, Bert and Bertrand, Alexander},
  booktitle={2025 47th Annual International Conference of the IEEE Engineering in Medicine and Biology Society (EMBC)}, 
  title={Temporal variation in the acoustic dynamic range is a confounding factor in {E}{E}{G}-based tracking of absolute auditory attention to speech}, 
  year={2025},
  volume={},
  number={},
  pages={1-4},
  doi={10.1109/EMBC58623.2025.11254180}}

@article{haupt2025neural,
  title={Neural Response Attenuates with Decreasing Inter-Onset Intervals Between Sounds in a Natural Soundscape},
  author={Haupt, Thorge and Rosenkranz, Marc and Bleichner, Martin G},
  journal={eneuro},
  volume={12},
  number={10},
  pages={ENEURO--0210},
  year={2025},
  publisher={Society for Neuroscience}
}

@misc{heintz2025postprocessingeegbasedauditoryattention,
      title={Post-processing of {E}{E}{G}-based Auditory Attention Decoding Decisions via Hidden Markov Models}, 
      author={Nicolas Heintz and Tom Francart and Alexander Bertrand},
      year={2025},
      eprint={2506.24024},
      archivePrefix={arXiv},
      primaryClass={eess.SP},
      url={https://arxiv.org/abs/2506.24024}, 
}

@article{fisher1915frequency,
  title={Frequency distribution of the values of the correlation coefficient in samples from an indefinitely large population},
  author={Fisher, R. A.},
  journal={Biometrika},
  volume={10},
  number={4},
  pages={507--521},
  year={1915}
}

@article{dempster1977maximum,
  title={Maximum likelihood from incomplete data via the EM algorithm},
  author={Dempster, Arthur P and Laird, Nan M and Rubin, Donald B},
  journal={Journal of the royal statistical society: series B (methodological)},
  volume={39},
  number={1},
  pages={1--22},
  year={1977},
  publisher={Wiley Online Library}
}

@article{smallwood2015science,
  title={The science of mind wandering: Empirically navigating the stream of consciousness},
  author={Smallwood, Jonathan and Schooler, Jonathan W},
  journal={Annual review of psychology},
  volume={66},
  number={1},
  pages={487--518},
  year={2015},
  publisher={Annual Reviews}
}

@article{warm2008vigilance,
  title={Vigilance requires hard mental work and is stressful},
  author={Warm, Joel S and Parasuraman, Raja and Matthews, Gerald},
  journal={Human factors},
  volume={50},
  number={3},
  pages={433--441},
  year={2008},
  publisher={SAGE Publications Sage CA: Los Angeles, CA}
}

@article{langner2013sustaining,
  title={Sustaining attention to simple tasks: a meta-analytic review of the neural mechanisms of vigilant attention.},
  author={Langner, Robert and Eickhoff, Simon B},
  journal={Psychological bulletin},
  volume={139},
  number={4},
  pages={870},
  year={2013},
  publisher={American Psychological Association}
}

@article{yao2024identifying,
  title={Identifying temporal correlations between natural single-shot videos and EEG signals},
  author={Yao, Yuanyuan and Stebner, Axel and Tuytelaars, Tinne and Geirnaert, Simon and Bertrand, Alexander},
  journal={Journal of Neural Engineering},
  volume={21},
  number={1},
  pages={016018},
  year={2024},
  publisher={IOP Publishing}
}

@inproceedings{xu2024convconcatnet,
  title={Convconcatnet: a deep convolutional neural network to reconstruct mel spectrogram from the eeg},
  author={Xu, Xiran and Wang, Bo and Yan, Yujie and Zhu, Haolin and Zhang, Zechen and Wu, Xihong and Chen, Jing},
  booktitle={2024 IEEE International Conference on Acoustics, Speech, and Signal Processing Workshops (ICASSPW)},
  pages={113--114},
  year={2024},
  organization={IEEE}
}

@article{accou2023decoding,
  title={Decoding of the speech envelope from EEG using the VLAAI deep neural network},
  author={Accou, Bernd and Vanthornhout, Jonas and hamme, Hugo Van and Francart, Tom},
  journal={Scientific Reports},
  volume={13},
  number={1},
  pages={812},
  year={2023},
  publisher={Nature Publishing Group UK London}
}

@article{lopez2025unsupervised,
  title={Unsupervised accuracy estimation for brain--computer interfaces based on selective auditory attention decoding},
  author={Lopez-Gordo, Miguel A and Geirnaert, Simon and Bertrand, Alexander},
  journal={IEEE Transactions on Biomedical Engineering},
  volume={72},
  number={8},
  pages={2388--2399},
  year={2025},
  publisher={IEEE}
}

@article{levy2025differential,
  title={Differential effects of external noise and situational interest on neurophysiological responses during video based learning},
  author={Levy, Orel and Shadi, Tal and Korisky, Adi and Bleichner, Martin G and Zion Golumbic, Elana},
  journal={npj Science of Learning},
  volume={10},
  number={1},
  pages={92},
  year={2025},
  publisher={Nature Publishing Group UK London}
}

@article{de2018decoding,
  title={Decoding the auditory brain with canonical component analysis},
  author={De Cheveign{\'e}, Alain and Wong, Daniel DE and Di Liberto, Giovanni M and Hjortkj{\ae}r, Jens and Slaney, Malcolm and Lalor, Edmund},
  journal={NeuroImage},
  volume={172},
  pages={206--216},
  year={2018},
  publisher={Elsevier}
}

@article{mann1947test,
  title={On a test of whether one of two random variables is stochastically larger than the other},
  author={Mann, Henry B and Whitney, Donald R},
  journal={The annals of mathematical statistics},
  pages={50--60},
  year={1947},
  publisher={JSTOR}
}

@article{mason2002areas,
  title={Areas beneath the relative operating characteristics ({ROC}) and relative operating levels ({ROL}) curves: Statistical significance and interpretation},
  author={Mason, Simon J and Graham, Nicholas E},
  journal={Quarterly Journal of the Royal Meteorological Society: A journal of the atmospheric sciences, applied meteorology and physical oceanography},
  volume={128},
  number={584},
  pages={2145--2166},
  year={2002},
  publisher={Wiley Online Library}
}

@article{benedek2014alpha,
  title={Alpha power increases in right parietal cortex reflects focused internal attention},
  author={Benedek, Mathias and Schickel, Rainer J and Jauk, Emanuel and Fink, Andreas and Neubauer, Aljoscha C},
  journal={Neuropsychologia},
  volume={56},
  pages={393--400},
  year={2014},
  publisher={Elsevier}
}

@article{geirnaert2026modeling,
  title={Modeling and Interpreting Correlations, Null Distributions and Significance Levels in Neural Tracking of Natural Stimuli},
  author={Geirnaert, Simon and Bertrand, Alexander and Francart, Tom and Vanthornhout, Jonas},
  journal={arXiv preprint arXiv:2608.10887},
  year={2026}
}

@misc{li2026multiscalegaussianmixturemodelinghmm,
      title={Multiscale {G}aussian-{M}ixture {M}odeling for {HMM} {P}ost-{P}rocessing in {S}elective {A}uditory {A}ttention {D}ecoding}, 
      author={Li, Tianyi and Geirnaert, Simon  and De Smedt, Bert  and Bertrand, Alexander},
      year={2026},
      eprint={2609.15490},
      archivePrefix={arXiv},
      primaryClass={eess.SP},
      url={https://arxiv.org/abs/2609.15490}, 
}

@article{heintz2025unsupervised,
  title={Unsupervised EEG-based decoding of absolute auditory attention with canonical correlation analysis},
  author={Heintz, Nicolas and Francart, Tom and Bertrand, Alexander},
  journal={arXiv preprint arXiv:2504.17724},
  year={2025}
}

@misc{icassp2027eegsaad,
  author       = {{EEG-sAAD-2027 Challenge Committee}},
  title        = {{ICASSP 2027 Grand Challenge of EEG-sAAD:
                   Selective Auditory Attention Decoding}},
  year         = {2026},
  howpublished = {Codabench},
  note         = {[Online]. Available:
                  \url{https://www.codabench.org/competitions/17856/}.
                  Accessed: Oct. 6, 2026}
}


\end{document}